# Solid state molecular rectifier based on self organized metalloproteins [**]

*By Ross Rinaldi[+], Adriana Biasco, Giuseppe Maruccio, Roberto Cingolani, Dario Alliata[•], Laura Andolfi, Paolo Facci, Francesca De Rienzo, Rosa Di Felice[*], and Elisa Molinari*

Recently, great attention has been paid to the possibility of implementing hybrid electronic devices exploiting the self-assembling properties of single molecules.[1] Impressive progress has been done in this field by using organic molecules[2,3] and macromolecules.[4,5] However, the use of biomolecules is of great interest because of their larger size (few nanometers) and of their intrinsic functional properties. Here, we show that electron-transfer proteins, such as the blue copper protein azurin (Az), can be used to fabricate biomolecular electronic devices exploiting their intrinsic redox properties, self assembly capability and surface charge distribution. The device implementation follows a bottom-up approach in which the self assembled protein layer interconnects nanoscale electrodes fabricated by electron beam lithography, and leads to efficient rectifying behavior at room temperature.

One of the fundamental goals of bioelectronics is the realization of nanoscale devices in which a few or a single biomolecule can be used to transfer and process an electronic signal. In order to design and realize such devices, several steps are required: *i)* choice and characterization of the suitable biomolecular system, *ii)* immobilization of the molecule onto an electronic substrate, *iii)* interconnection to contacts, *iv)* device fabrication, *v)* read-out, and *vi)* processing of the information. The nano bio-technology community has started a big effort along these lines to realize biomolecular devices for information technology.[6] Among biomolecules, proteins have a fundamental role in biological processes. The combination of molecular biology (for engineering

[**] Support for this research was provided by the Italian Istituto Nazionale per la Fisica della Materia (PRA-SINPROT for funding and Commissione Calcolo for computing time at CINECA) and by the EU (SAMBA Project). The authors gratefully acknowledge Prof. G. W. Canters for lively and fruitful discussions.

[+] Prof. R. Rinaldi, A. Biasco, G. Maruccio, Prof. R. Cingolani, National Nanotechnology Laboratory of INFM, Dipartimento di Ingegneria dell'Innovazione, Università di Lecce, Lecce, Italy
[•] Dr. D. Alliata, Dr. L. Andolfi, INFM Viterbo, Università della Tuscia, Viterbo, Italy
[*] Dr. R. Di Felice, Dr. F. De Rienzo, Prof. E. Molinari, Dr. P. Facci, INFM Center on nanoStructures and bioSystems at Surfaces (S³), Dipartimento di Fisica, Università di Modena e Reggio Emilia, Via Campi 213/A, 41100 Modena, Italy

proteins with the desired functional and/or self-assembling properties) and nanotechnology (for device fabrication) thus becomes the tool to realize a new class of nano-electronic elements.[7]

Thanks to their functional characteristics, metalloproteins appear to be good candidates for biomolecular nanoelectronics. Among them, blue copper proteins,[8] and in particular azurin, are candidates of choice, due to some specific structural properties and intrinsic functionality in biological environments. They can bind gold via a disulfide site present on the protein surface, and their natural electron transfer activity can be exploited for the realization of molecular switches whose conduction state can be controlled by tuning their redox state through an external voltage source (gate). Recent experiments carried out by electrochemical scanning tunneling microscopy (STM) on Az molecules immobilized on Au(111) surfaces have demonstrated the possibility of eliciting current flow through the redox level of a single molecule, providing the physical ground for using these metalloproteins as molecular switches.[9,10] However, in order to verify the possibility of realizing a real biomolecular device in the solid state and operating in air, a very precise study of the redox, electronic and electrical properties of the metalloproteins linked to an inorganic substrate under non-physiological environments is required.

We show here that a biomolecular electron rectifier in the solid state can be implemented by interconnecting an azurin monolayer immobilized on $SiO_2$ with two gold nanoelectrodes. This rectifier exhibits discrete current steps in the positive wing of the I-V curve, which are ascribed to resonant tunneling through the redox active center. This result represents a fundamental step for the realization of protein-based electronic devices, such as field effect transistors (FET) or single electron transistors (SET).

Azurin[8] from *Pseudomonas aeruginosa* is a small (molecular mass 14.6 kDa, Fig. 1(a)) and soluble metalloprotein involved in the respiratory phosphorylation of its hosting bacterium. Structural[11] and electronic[12] studies have shown that the Az capability to function as a one-electron carrier, in the biological environment, is due the equilibrium between the two stable oxidation states of the Cu ion, $Cu^{+1}$ (reduced) and $Cu^{+2}$ (oxidized), and to the structural stability of the active site. The redox active centre of azurin contains a copper ion liganded to 5 aminoacid atoms in a peculiar ligand-field symmetry, which endows the protein with unusual spectroscopic and electrochemical properties such as an intense absorption band at 628 nm, a small hyperfine splitting in the electron paramagnetic spectrum,[13] and an unusually large equilibrium potential (+116 mV *vs* SCE)[14] in comparison to the Cu(II/I) aqua couple (-89 mV *vs* SCE).[15] The absorption spectrum of an azurin solution[16] is reported in Fig. 1(b). The broad band centred at 628 nm is related to S(Cys)→Cu charge-transfer transitions,[17] while the peak centred at 275 nm is due to electronic transitions originated from UV light absorption of aromatic residues of the protein.

While the redox properties of azurin can be exploited for obtaining a current flux, its peculiar structural properties can be exploited for immobilizing the protein onto substrates. The surface disulfide bridge Cys3-Cys26 (Fig. 1(a)) may be used to bind the protein to gold (or other electronically soft metals), thus envisaging the possibility to deposit oriented layers on gold substrates.[9,10,18] The achievement of oriented immobilization is extremely important for electronic applications in which charge transport benefits of the long range order of the transporting material. Orientation can in principle affect conduction in two ways: (i) it can allow increased protein coverage, thus favoring electron transfer among neighboring molecules sitting at closer distances, or (ii) enhance, for a given coverage, intermolecular electron transfer (due to the fact that the positions of the Cu-sites are approximately coplanar, thus offering more favorable pathways for conduction). Another viable orientation mechanism is associated to the role played by electrostatics in solution, which drives the protein self-assembly onto the substrate, and the resulting macroscopic electric fields that may be present in the device.

To gain insight into this last mechanism, we examined the molecular electrostatic potential (MEP) of the protein in solution. MEPs are known to be especially important in protein interaction properties at medium and long-range distances in solution, and play a fundamental role in the recognition processes of biomolecules.[19] It is therefore expected that electrostatics will influence both the deposition kinetics of the proteins in solution, and their self-assembly. The charge distribution on the azurin surface gives origin to an electric dipole, as clearly highlighted in Fig. 1(c), where the azurin MEP is reported. The occurrence of such a strong intrinsic dipole (150 Debye) suggests that two-terminal circuits, interconnecting solid state films of immobilized azurin molecules, have a resultant polarity dependent on the value and on the orientation of the individual molecular dipoles. Both oriented and non oriented azurin films should exhibit a macroscopic polarization. However, the total dipole may be enhanced by depositing oriented self-assembled films with parallel dipoles, that would sum up along their common direction. If this dipole distribution is preserved in the devices after drying, we expect that it will induce a macroscopic electric field favoring conduction, thus influencing the electronic characteristics of the device.

The planar biomolecular devices studied in this work consist of two Au nanoelectrodes separated by a small gap, where a solid state azurin monolayer was immobilized. Random and oriented immobilizations[20-22] were attained according to the procedures reported in the experimental section. Different experimental techniques were used to characterize the immobilized protein layers. The formation of extended protein monolayers onto oxygen exposing surfaces was tested by SFM imaging both at low and molecular resolution (see below). The thickness of the layer measured by engraving it with the SFM tip matches fairly well the expected value for the supramolecular architecture (about 4 nm). X-ray photoemission spectroscopy (XPS) performed at different stages of the sample preparation confirmed the presence of the expected elements and the correct growth of the film.[20]

Fig.1(d) shows the AFM topographical image of an Az film deposited onto a $SiO_2$ substrate. Images have been acquired in non-contact mode. The lateral size of the visible features ranges around 12 nm, due to tip-sample convolution (the actual size of Az from X-ray crystallography[23] is about 4 nm). The quality of the pictures is further influenced by substrate roughness, improving a lot on atomically flat substrates.[20] The height of the visible features is instead about 4 nm, consistent with single protein size. No strong differences in the surface morphology of uniformly oriented and randomly oriented protein layers was found,[24] but their difference has been tested by thorough XPS investigations on the Cu levels. Photocurrent experiments on sample A were performed in order to understand the electronic structure of the protein in the solid state and to compare the results with the absorption spectrum of the protein in solution. This analysis is important, because we would like our novel devices to operate in air and not in the aqueous environment occurring in living cells. This is a different approach to the fascinating one aimed at implementing computations with macromolecules (*e.g.*, DNA[25]), living cells (*e.g.*, neurons[26]), or small organisms (*e.g.*, bacteria[27]) in the liquid environment, which might also lead to impressive results in the next decade. In Fig. 1(b), the photocurrent spectrum measured on a solid state Az layer by biasing the sample at 5 V is superimposed to the absorption spectrum of Az in solution. The photocurrent exhibits a strong absorption onset for wavelengths smaller than 400 nm, consistent with the absorption spectrum. In addition, the absorption band around 628 nm observed in the azurin solution spectrum corresponds to a clear current minimum in the photocurrent spectrum. This effect indicates that no charge transfer occurs through the Cu atom when it is virtually reduced, and suggests that the electron transfer mechanism through the Az redox site can be exploited to implement functional electronic and optoelectronic devices.

Current-voltage measurements were performed on all samples soon after deposition and at different delays after the deposition process, under the same experimental conditions (room temperature, atmospheric pressure, 54% of humidity). In Fig. 2, we show the comparison between the current-voltage curves measured in the oriented (sample B) and randomly oriented (sample A) azurin layers. The continuous and dotted lines represent the downward and upward sweeps, respectively. Three important effects can be deduced from this comparison.

*Asymmetry*. Both curves are asymmetric, with a strong rectifying behavior. The value of the current measured under forward bias between the nanoelectrodes suggests that the electron transfer mechanisms in the protein, between the Cu site and the edges, is quite effective in determining the conduction processes. The difference between the positive and negative wings of the current curve of both sample A and B may be attributed to the presence of the dipole in the azurin molecules which sets the polarization of the planar devices. The rectification ratio at 1.5 V amounts to 175 for the devices based on the oriented layers and to 10 for the non-oriented ones. At larger bias the difference reduces, because the reverse current in the devices with the oriented layers increases.

*Current Intensity.* The current flowing through the device with the oriented layer is about ten times larger than that flowing through the device with the non-oriented layer. In fact, the regular orientation of the Az molecules in sample B, determined by the unique sticking site on the protein (the Cys3-Cys26 bridge) exploited for the layer formation, drives the self-assembly on the substrate, resulting in a distribution of parallel dipoles, as schematically depicted in Fig. 2. This induces a macroscopic electric field favoring conduction. In sample A, although electrostatic long range interaction will anyway drive the kinetics of protein deposition thus favoring a dipole to dipole surface organization of the proteins, a complete parallel orientation of the molecular dipoles cannot be achieved as a consequence of the many possible sticking sites on the protein surface used in the immobilization procedure. This favors the randomization of the dipole orientation and statistically reduces the total dipole field. It is also worth noting that in the oriented protein layer intermolecular electron transfer can be further enhanced by the spatial arrangement of the Cu active sites which turn out to be approximately coplanar, thus offering more favorable pathways for conduction.

*Steps.* Distinct steps are observed at voltages which depend on the orientation of the proteins. We observe that in the oriented protein layer under forward bias the current is step-like with a smooth exponential rise in the region between 1.9 V and 2.3 V, and a steep rise around 4.9 V. The step centered around 2.1 V corresponds to the energy required by the protein molecule to reduce the Cu atom by means of the electronic transition involving the S(Cys)→Cu charge-transfer. This corresponds to the 628 nm band in the absorption spectrum shown in Fig.1(b). The step around 4.9 V corresponds to the energy required to perform resonant tunneling via the redox levels of azurin.[9] Such a process occurs through a coherent two-step tunneling in which the electrons go from the negative to the positive electrode via the molecular redox level. This is consistent with the electrochemical STM measurements and in situ cyclic voltammetry curves, showing a maximum at

–4.96eV (measured with respect to the vacuum level).[9] This step is also observable in the I-V curve of sample A, though less pronounced and with a small hysteresis loop (in the upward voltage sweep the step is registered at 4.9 V, while in the downward sweep there are two steps, at 2.4 V and at 7.6 V). The 4.9 V step is observed in all the fabricated samples, both in the upward and downward sweeps, and regardless of the orientation of molecules, suggesting the intrinsic molecular origin of the effect. A second step at 9.1 V is also visible in the I-V curve of sample B, indicating the existence of a double (sequential) electron transfer process through the redox level of the molecule.[10] The shape of the I-V curve in the voltage range between the two steps does not exhibit a monotonic behavior: its fine structure might be associated to the intrinsic charge transport mechanisms, but an accurate interpretation is beyond the scope of the present report and is the subject of ongoing theoretical investigation.

Measurements performed after one day and one week of delay gave lower current intensities and changes in the value of the zero current region, indicating the sensitivity of the metalloprotein to humidity and temperature variations, that could lead to protein unfolding and/or Cu reduction. Such issues compromise the stability of device operation in air, and must be addressed in order to attain reliable devices. Finally, similar I-V measurements performed on the reference samples containing just the 3-APTS, 3-APTS+GD, and 3-MPTS molecular layers (see the experimental description), showed current values close to the noise level of the electrometer with strong oscillations. Therefore, no contribution to the overall transport in the device is expected to come from the supramolecular layers used to prepare the surface to the protein linking.

In conclusion, we have demonstrated the first electronic device based on proteins in the solid state operating in air. All the fabricated devices are rectifiers with a step-like current under forward bias. The photocurrent spectrum measured on the devices indicates that the electron transfer mechanism through the Az redox site can be exploited to implement three terminal electronic and optoelectronic devices. A strong role in the conduction mechanisms seems to be played by the charge distribution of the molecular layer into the implemented solid state devices.

*Experimental Details*

Commercial azurin from *Pseudomonas aeruginosa* (Sigma) was used without further purification after having checked that the ratio $OD_{628}/OD_{280}$ ($OD_\lambda$ = optical density measured at λ nm) was in accordance with the literature values (0.53 - 0.58).[10] A working solution of $10^{-4}$ M azurin in 50 mM $NH_4Ac$ (Sigma) buffer, pH 4.6, was prepared. The buffer was degased with $N_2$ flow prior to use. Milli-Q grade water (resistivity 18.2 MΩcm) was used throughout all the experiments.

The nanoelectrodes have an arrow shape, are separated by a gap of 70 nm, and were fabricated by electron beam lithography and lift-off on a $Si/SiO_2$ substrate. Two different immobilization procedures were developed for the azurin on $Si/SiO_2$, exploiting different structural features of the protein and resulting in (1) random orientation (3-step procedure) and (2) oriented immobilization (2-step procedure) in the gap of the planar circuit. Following the 3-step procedure (1), the immobilization of the proteins is achieved through the formation of imido-bonds between the exposed amino groups on the azurin surface and the carbonyl heads exposed on the functionalized $SiO_2$ substrate. This procedure is based on a 3-step chemical reaction:[20] first, the $Si/SiO_2$ substrates were incubated with gold electrodes in 3-aminopropyltriethoxysilane (3-APTS) for 2 minutes and rinsed in $CHCl_3$, in order to remove 3-APTS molecules not linked to the surface; second, the sample already reacted with silanes was exposed to glutharic dialdheyde GD for 10 minutes and then thoroughly washed in ultra-pure $H_2O$; third, the pre-coated substrates were exposed to the azurin solution for 10 minutes and rinsed in $NH_4Ac$ to get rid of physisorbed molecules (sample A). Because the azurin molecule contains 13 surface-exposed amino groups (found in the Lys residues and in the N-termini), random orientation of the proteins is obtained in the solid state by this immobilization method. In contrast, a highly ordered phase is achieved following procedure (2), that exploits a 2-step immobilization protocol acting on the unique disulfide bond found in the protein between Cys3 and Cys26.[21] Such a procedure not only provides a special orientation of the protein molecules, but also facilitates the chemisorption process by avoiding the GD exposure. In this case, $Si/SiO_2$ substrates with the gold electrodes were incubated with 3-mercaptopropyltrimethoxysilane (3-MPTS) for 2 min, rinsed in abundant $CHCl_3$, then exposed to the azurin solution for typically 40-60 minutes, and thoroughly rinsed in pure buffer (sample B). Protein immobilization took place via the reaction of the free thiol groups of 3-MPTS with the surface disulfide bridge of Az, giving rise to substrate/overlayer disulfide bonds. By using the 2-step method for the immobilization, the orientation of the molecule is expected to be well defined, as confirmed by electrochemical (cyclic voltammetry) and SFM measurements.[21] The disulfide bridge is located, in the barrel-like shape of the azurin, at the opposite site of the Cu ion. The sulphur atom of the Cu-ligand Cys112 could in principle be involved in binding the substrate. However, this mechanism would lead to loss of the copper atom and destruction of the redox site architecture, thus altering the redox nature of the protein. STM/STS studies on azurin molecules bond to Au surface give in fact indication that the last mechanism is highly improbable.[22]

Samples covered just by the first-stage molecular layers in both immobilization procedures (namely 3-MPTS, 3-APTS and 3-APTS+GD) were also produced for reference, to evaluate the impact of each processing step on the final layer characteristics. 3-APTS (Sigma) and 3-MPTS were diluted to 6.6% (V/V) in $CHCl_3$ immediately prior to use. GD (Sigma) was diluted in $H_2O$ (Milli-Q grade) to a final concentration of 4 $10^{-4}$ M.

*Theory*

The MEP of azurin from *Pseudomonas aeruginosa* was computed on the X-ray oxidized protein structure obtained at pH 5.5 (structure A in the PDB file 4azu, http://pdb.gmd.de). The calculation was performed with the University of Houston Brownian Dynamics (UHBD) software[28], as described by De Rienzo et al.[19]. The MEP was computed at an ionic strength of 50 mM, in order to simulate the experimental conditions used for protein deposition onto the functionalized Si/$SiO_2$ substrate. The N- and C-terminal residues were charged. The ionizable residues (glu, asp, lys, arg) were in their usual protonation states at pH 5.5. The histidines 35 and 83 were protonated, due to their pKa values[29]. The calculated molecular electric dipole is of the order of 150 Debye. It should be pointed out that this value was obtained for a frozen structure: in solution, the overall charge distribution on the protein surface and the dipole moment continually fluctuate, and the actual dipole is averaged over many frozen configurations. To verify that the presence of such a dipole is not an artifact due to the particular frozen conformation of azurin in the crystal, the MEP was also calculated for a relaxed equilibrium geometry (CHARMM simulation[30]): the resulting dipole did not differ significantly from that of the X-ray structure.

**Figure Captions.**

Figure 1.

a) Structure of azurin from *Pseudomonas Aeruginosa*. -b) Absorption spectrum of azurin in solution (dashed line) and photocurrent spectrum (continuous line) from an immobilized solid state film of azurin on Si/SiO$_2$ interconecting two Au nanocontacts (sample A, discussed in the text). The photocurrent measurements were carried out under a bias voltage of 5 V. The Halogen lamp used for the photocurrent could not reach the wavelength range around 275 nm, probed by the absorption spectrophotometer. c) MEP of the oxidized azurin. Iso-potential surfaces are shown at -0.5 (red) and 0.5 kcal/mol/e (blue). The Cu site is identified with a white circle, and the yellow frame indicates the Cys3-Cys26 disulfide bridge.– d) AFM image of an azurin monolayer. The AFM measurements were performed in non-contact mode and in air on one of the deposited monolayer in a working device (CP research system THERMOMICROSCOPE). The typical tip-sample interaction force was 1 nN at 1 Hz scan rate.

Figure 2.

Current voltage curves of samples A and B. The gap between the gold electrodes was 70 nm. Current-voltage characteristics of the samples were measured by using a high sensitivity electrometer (Keithly 6517) with sub picoAmpere noise level. All the measurements were performed after drying the layer under Nitrogen flux. The possible arrangement of proteins between the electrodes in the oriented (top-left) and non-oriented (bottom-right) layers are illustrated schematically . The positive and negative potential regions, the Cu site and the disulfide bridge are indicated by the same color code as in Fig.1(c).

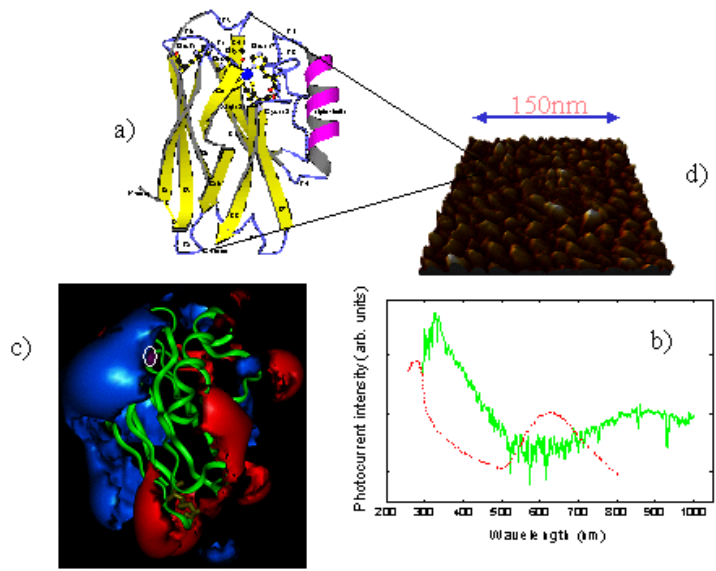

**Figure 1.**

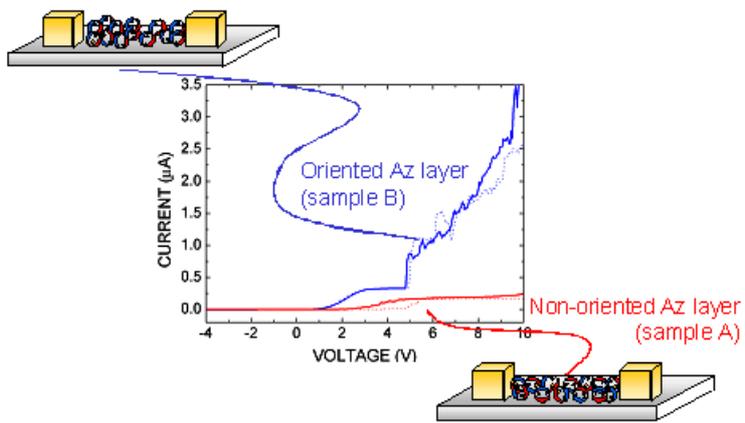

**Figure 2.**